\documentclass[10pt]{revtex4}
\usepackage{graphicx}

\newcount\driver
\newcount\bozza

\font\ottorm=cmr8\font\ottoi=cmmi8\font\ottosy=cmsy8%
\font\ottocss=cmcsc8%
\font\sixrm=cmr6\font\sixi=cmmi6\font\sixsy=cmsy6%
\font\fiverm=cmr5\font\fivesy=cmsy5
\font\fivei=cmmi5
\font\tenmib=cmmib10
\font\sevenmib=cmmib10 scaled 800

 2

\font\cs=cmcsc10
\font\sc=cmcsc10

\font\elevenrm=cmr11
\font\twelverm=cmr12
\font\ottorm=cmr8
\font\msytw=msbm9 scaled\magstep1
\font\msytww=msbm7 scaled\magstep1
\font\msytwww=msbm5 scaled\magstep1
\font\indbf=cmbx10 scaled\magstep2

\font\ottorm=cmr8\font\ottoi=cmmi8\font\ottosy=cmsy8%
\font\ottocss=cmcsc8%
\font\sixrm=cmr6\font\sixi=cmmi6\font\sixsy=cmsy6%
\font\fiverm=cmr5\font\fivesy=cmsy5
\font\fivei=cmmi5

\def\ottopunti{\def\rm{\fam0\ottorm}%
\textfont0=\ottorm\scriptfont0=\sixrm\scriptscriptfont0=\fiverm%
\textfont1=\ottoi\scriptfont1=\sixi\scriptscriptfont1=\fivei%
\textfont2=\ottosy\scriptfont2=\sixsy\scriptscriptfont2=\fivesy%
\textfont4=\ottocss\scriptfont4=\sc\scriptscriptfont4=\sc%
\scriptfont4=\ottocss\scriptscriptfont4=\ottocss%
\textfont5=\tenmib\scriptfont5=\sevenmib\scriptscriptfont5=\fivei
\setbox\strutbox=\hbox{\vrule height7pt depth2pt width0pt}%
\normalbaselineskip=9pt\let\sc=\sixrm\normalbaselines\rm}

\mathchardef\BDpr = "0540  %Dpr
\mathchardef\Bg   = "050D  %gamma

{\count255=\time\divide\count255 by 60 \xdef\hourmin{\number\count255}
        \multiply\count255 by-60\advance\count255 by\time
   \xdef\hourmin{\hourmin:\ifnum\count255<10 0\fi\the\count255}}

\def\openone{\leavevmode\hbox{\elevenrm 1\kern-3.63pt\twelverm1}}%
\def\*{\vglue0.5truecm}

\let\a=\alpha \let\b=\beta  \let\g=\gamma  \let\d=\delta \let\e=\varepsilon
     \let\th=\theta  \let\l=\lambda
\let\m=\mu    \let\n=\nu             \let\r=\rho
 \let\t=\tau   \let\f=\varphi 
   
 \let\D=\Delta

\def\\{\hfill\break} \let\==\equiv

\let\io=\infty 

\let\0=\noindent

\def\media#1{{\langle#1\rangle}}
\def\ie{{\it i.e.}}

\def\tende#1{\,\vtop{\ialign{##\crcr\rightarrowfill\crcr
 \noalign{\kern-1pt\nointerlineskip} \hskip3.pt${\scriptstyle
 #1}$\hskip3.pt\crcr}}\,}
\def\circage{\lower2pt\hbox{$\,\buildrel > \over {\scriptstyle \sim}\,$}}
\def\otto{\,{\kern-1.truept\leftarrow\kern-5.truept\to\kern-1.truept}\,}

\def\EE{{\cal E}} \def\VV{{\cal V}}
\def\CC{{\cal C}}\def\FF{{\cal F}}

\def\T#1{{#1_{\kern-3pt\lower7pt\hbox{$\widetilde{}$}}\kern3pt}}
\def\VVV#1{{\VV #1}_{\kern-3pt
\lower7pt\hbox{$\widetilde{}$}}\kern3pt\,}
\def\W#1{#1_{\kern-3pt\lower7.5pt\hbox{$\widetilde{}$}}\kern2pt\,}

\def\indica{\leaders \hbox to 0.5cm{\hss.\hss}\hfill}
\def\guida{\leaders\hbox to 1em{\hss.\hss}\hfill}

\def\hhh{{\bf h}}

\def\VV#1{{\,\underline#1\,}}

\mathchardef\aa   = "050B
\mathchardef\bb   = "050C
\mathchardef\ggg  = "050D
\mathchardef\xxx  = "0518
\mathchardef\zzzzz= "0510
\mathchardef\oo   = "0521
\mathchardef\lll  = "0515
\mathchardef\Dp   = "0540
\mathchardef\H    = "0548
\mathchardef\FFF  = "0546
\mathchardef\ppp  = "0570
\mathchardef\Bn   = "0517
\mathchardef\pps  = "0520
\mathchardef\fff  = "0527
\mathchardef\FFF  = "0508
\mathchardef\nnnnn= "056E

\def\to{\rightarrow}

\def\qed{\hfill\raise1pt\hbox{\vrule height5pt width5pt depth0pt}}

\def\indic{\hbox{\raise-2pt \hbox{\indbf 1}}}

\def\RRR{\hbox{\msytw R}} \def\rrrr{\hbox{\msytww R}}
\def\rrr{\hbox{\msytwww R}} 
 
\def\NNN{\hbox{\msytw N}} 
 \def\ZZZ{\hbox{\msytw Z}}
 \def\zzz{\hbox{\msytwww Z}}

\def\V0{{\bf 0}}

\font\tenmib=cmmib10 
\font\sevenmib=cmmib7\font\fivemib=cmmib5 

\font\fivei=cmmi5\font\sixi=cmmi6\font\ottoi=cmmi8
\font\ottorm=cmr8\font\fiverm=cmr5\font\sixrm=cmr6
\font\ottosy=cmsy8\font\sixsy=cmsy6\font\fivesy=cmsy5%%
\font\ottocss=cmcsc8%

\mathchardef\Ba   = "050B  %alfa
\mathchardef\Bb   = "050C  %beta
\mathchardef\Bg   = "050D  %gamma
\mathchardef\Bd   = "050E  %delta
\mathchardef\Be   = "0522  %varepsilon
\mathchardef\Bee  = "050F  %epsilon
\mathchardef\Bz   = "0510  %zeta
\mathchardef\Bh   = "0511  %eta
\mathchardef\Bthh = "0512  %teta
\mathchardef\Bth  = "0523  %varteta
\mathchardef\Bi   = "0513  %iota
\mathchardef\Bk   = "0514  %kappa
\mathchardef\Bl   = "0515  %lambda
\mathchardef\Bm   = "0516  %mu
\mathchardef\Bn   = "0517  %nu
\mathchardef\Bx   = "0518  %xi
\mathchardef\Bom  = "0530  %omi
\mathchardef\Bp   = "0519  %pi
\mathchardef\Br   = "0525  %ro
\mathchardef\Bro  = "051A  %varrho
\mathchardef\Bs   = "051B  %sigma
\mathchardef\Bsi  = "0526  %varsigma
\mathchardef\Bt   = "051C  %tau
\mathchardef\Bu   = "051D  %upsilon
\mathchardef\Bf   = "0527  %phi
\mathchardef\Bff  = "051E  %varphi
\mathchardef\Bch  = "051F  %chi
\mathchardef\Bps  = "0520  %psi
\mathchardef\Bo   = "0521  %omega
\mathchardef\Bome = "0524  %varomega
\mathchardef\BG   = "0500  %Gamma
\mathchardef\BD   = "0501  %Delta
\mathchardef\BTh  = "0502  %Theta
\mathchardef\BL   = "0503  %Lambda
\mathchardef\BX   = "0504  %Xi
\mathchardef\BP   = "0505  %Pi
\mathchardef\BS   = "0506  %Sigma
\mathchardef\BU   = "0507  %Upsilon
\mathchardef\BF   = "0508  %Fi
\mathchardef\BPs  = "0509  %Psi
\mathchardef\BO   = "050A  %Omega
\mathchardef\BDpr = "0540  %Dpr
\mathchardef\Bstl = "053F  %*

\def\V#1{{\bf#1}}
\let\aa=\Ba\let\fff=\Bf
\let\oo=\Bo\let\nn=\Bn
\let\pps=\Bps\def\hhh={\V h}
\let\bb=\Bb

\def\RRR{\hbox{\msytw R}} \def\rrrr{\hbox{\msytww R}}
\def\rrr{\hbox{\msytwww R}} 
 
\def\NNN{\hbox{\msytw N}} 
 \def\ZZZ{\hbox{\msytw Z}}
 \def\zzz{\hbox{\msytwww Z}}

\def\ins#1#2#3{\vbox to0pt{\kern-#2 \hbox{\kern#1 #3}\vss}\nointerlineskip}
\newdimen\xshift \newdimen\xwidth \newdimen\yshift
\newcount\griglia

\def\insertplot#1#2#3#4#5#6{%
\begin{figure}[h]
\begin{center}
\vspace{#2pt}
\begin{minipage}{#1pt}
#3
\ifnum\driver=1
\griglia=#6
\ifnum\griglia=1
\openout13=griglia.ps
\write13{gsave .2 setlinewidth}
\write13{0 10 #1 {dup 0 moveto #2 lineto } for}
\write13{0 10 #2 {dup 0 exch moveto #1 exch lineto } for}
\write13{stroke}
\write13{.5 setlinewidth}
\write13{0 50 #1 {dup 0 moveto #2 lineto } for}
\write13{0 50 #2 {dup 0 exch moveto #1 exch lineto } for}
\write13{stroke grestore}
\closeout13
\includegraphics{griglia.ps}\fi
\includegraphics{#4.ps}\fi
\ifnum\driver=2
\fi
\end{minipage}
\end{center}
\caption{#5}
\end{figure}
}

\newdimen\shift \shift=-1truecm
\def\lb#1{%
\ifnum\bozza=1
\label{#1}\rlap{\kern\shift{$\scriptstyle#1$}}
\else\label{#1}
\fi}

\def\be{\begin{equation}}
\def\ee{\end{equation}}
\def\bea{\begin{eqnarray}}\def\eea{\end{eqnarray}}
\def\bean{\begin{eqnarray*}}\def\eean{\end{eqnarray*}}
\def\bfr{\begin{flushright}}\def\efr{\end{flushright}}
\def\bc{\begin{center}}\def\ec{\end{center}}
\def\ba#1{\begin{array}{#1}} \def\ea{\end{array}}
\def\bd{\begin{description}}\def\ed{\end{description}}
\def\bv{\begin{verbatim}}\def\ev{\end{verbatim}}
\def\nn{\nonumber}
\def\Halmos{\hfill\vrule height10pt width4pt depth2pt \par\hbox to \hsize{}}

\newdimen\xshift \newdimen\xwidth \newdimen\yshift \newdimen\ywidth

\def\ins#1#2#3{\vbox to0pt{\kern-#2\hbox{\kern#1 #3}\vss}\nointerlineskip}

\def\eqfig#1#2#3#4#5{
\par\xwidth=#1 \xshift=\hsize \advance\xshift
by-\xwidth \divide\xshift by 2
\yshift=#2 \divide\yshift by 2
\line{\hglue\xshift \vbox to #2{\vfil
#3 \includegraphics{#4.ps}
}\hfill\raise\yshift\hbox{#5}}}

\def\8{\write12}

\begin{document}

\title{Periodic minimizers in 1D local mean field theory}

\author{Alessandro Giuliani}\thanks{ Present Address Dipartimento di 
Matematica di Roma Tre, Largo S. Leonardo Murialdo 1, 00146 Roma, Italy.\\
\\
\copyright\, 2007  by the authors. This paper may be
reproduced, in its
entirety, for non-commercial purposes.}
\affiliation{Department of Physics, Princeton University, Princeton 08544 NJ, 
USA.}
\author{Joel L. Lebowitz}
\affiliation{Departments of Mathematics and Physics, Rutgers University,
Piscataway, NJ 08854 USA.}
\author{Elliott H. Lieb}
\affiliation{Departments of Mathematics and Physics, 
Princeton University, Princeton, NJ 08544 USA.}
\vspace{3.truecm}
\date{December 13, 2007}%  version 13}
\begin{abstract} Using reflection positivity techniques we prove the
existence of minimizers for a class of mesoscopic free-energies 
representing 1D systems with competing interactions. 
{\it All} minimizers are either periodic, with zero average, or of
constant sign. If the local term in the free energy satisfies a convexity 
condition, then all minimizers are either periodic or constant.
Examples of both phenomena are given. This extends our previous
work where such results were proved for the ground states of lattice
systems with ferromagnetic nearest neighbor interactions and dipolar
type antiferromagnetic long range interactions.   
\end{abstract}

\maketitle

\renewcommand{\thesection}{\arabic{section}}

%%%%%%%%%%%%%%%%%%%%%%%%%%%%%%%%%%%%%%%%%%%%%%%%%%%%%%%%%%%%%%%%%%%%%%%%%%
%%%%%%%%%%%%%%%%%%%%%%%%%%%%%%%%%%%%%%%%%%%%%%%%%%%%%%%%%%%%%%%%%%%%%%%%%%
\section{Introduction}
\setcounter{equation}{0}
%%%%%%%%%%%%%%%%%%%%%%%%%%%%%%%%%%%%%%%%%%%%%%%%%%%%%%%%%%%%%%%%%%%%%%%%%%
%%%%%%%%%%%%%%%%%%%%%%%%%%%%%%%%%%%%%%%%%%%%%%%%%%%%%%%%%%%%%%%%%%%%%%%%%%

We consider the nature of the minimizers for a class of 1D free-energy
functionals that model the continuum limit of microscopic
systems with competing interactions on different length scales. An
example is a spin system on a lattice with a nearest neighbor
ferromagnetic interaction and a long range antiferromagnetic power law
type interaction. In \cite{GLL06} we considered the ground states of
such systems in one-dimension and in \cite{GLL07} we also investigated
higher dimensional models with dipolar type interactions. 
In both cases we obtained periodic ground
states whose period (which could be infinite) depended on both the
strength of the short range interaction and the nature of the long
range interaction. The technique used in those papers, reflection
positivity, could not be extended to positive temperatures,
for which only approximate methods and computer simulations are
available now \cite{MWRD95,AWMD96,GTV00,SS00}. 
It turns out, however, that these reflection positivity
methods are directly applicable to the Ginzburg-Landau type free-energy 
functionals used to describe the continuum versions of such microscopic 
systems \cite{M,SA}. These include
finite temperature effects, at least at a mean field level, 
via an inclusion of a local entropy term in the effective continuum description
of the system.  

These free-energies functionals have been used extensively in
both the physical and mathematical literatures as models for a great
variety of systems, including micromagnets \cite{Br,DKMO,GD82}, 
diblock copolymers \cite{HS,L80,OK},
anisotropic electron gases \cite{SK04,SK06}, 
polyelectrolytes \cite{BE88}, charge-density waves in 
layered transition metals \cite{Mc75} and superconducting films 
\cite{EK93}.
Many of these systems are characterized by
low temperature phases displaying spontaneous formation of periodic
mesoscopic patterns, such as stripes or bubbles.  It is, therefore,
important to show that these free energy functionals have periodic minimizers. 
This has been proved rigorously in some cases \cite{AM,CO} and argued for 
heuristically in others \cite{Br,BL,EK93,GD82,GP3,HS,L80,M,SK04,SK06}.

%However, a full microscopic theory of these phenomena is still lacking, 
%even at 
%zero temperature.  In fact, most microscopic  studies are based
%on periodicity assumptions that are very hard to prove. As far as we
%know, there are just a few cases where periodicity of the ground state of such 
%systems
%have been proved. 

In this paper we use reflection positivity methods to prove the periodicity of
minimizers for a certain class of such 1D free-energy functionals. These
include cases that were not treated before, e.g., those with
long-range power law type interactions. As noted before, reflection 
positivity methods have been
succesfully applied to find periodic ground states for a class of microscopic 
1D and 2D lattice spin models but has not, as far as we know, been used 
before for continuum systems. 
We begin in Section \ref{two} by presenting the class of
models under consideration and our results. These are proved in
Section \ref{three}. In Section \ref{four} we give an example of the
transition from a state of finite periodicity to a uniform (infinite
periodicity) state, as a parameter is varied.  We discuss the
connection with related work in Section \ref{five}. 

\section{Formulation of model and statement of results}     \label{two}

The formal infinite volume free energy functional to be minimized, 
in a sense to be made precise below, is
\be \label{energy}
\EE(\phi)=\int_{\rrr} dx\, \Big[\big(\phi'(x)\big)^2+F(\phi)\Big]+
\int_{\rrr}dx \int_{\rrr} dy\; \phi(x) v(x-y) \phi(y)
\;,\qquad v(x)=\l\int_0^\io \n(d\a)\, e^{-|x|\a}\;,
\label{1.1}\ee
with $\n(d\a)$ a probability measure such that
$\l\int_0^\io\n(d\a)\a^{-1}=\int_0^\io v(x)\,dx <+\io$, 
$\l$ a positive constant. We shall
also assume that $F(t)$ is an even function of its argument,
and that $F(t)\geq 0$, 
with $F(t)>0$ for $|t|<1$ and $F(t)=0$ for $|t|=1$. Note that we do not need 
either that $F$ is continuous or that it goes to $+\io$ as $|t|\to \io$.
Some examples to keep in mind are $F(t)=(t^2-1)^{2}$ or $F(t)=(|t|-1)^{2}$ or
$F(t)=a(t)-a(1)$, where, defining $\a=\tanh\b$:  
\be a(t)=\left\{\matrix{-t^2+(\a \b)^{-1}\big[(1+\a t)\log(1+\a t)+(1-\a t)
\log(1-\a t)\big]\;, \ \ {\rm if}\ \ |t|<\a^{-1}\;,\cr +\io\;,\hfill 
\ \ {\rm if}\ \ |t|\ge\a^{-1}\;. \cr}\right.\ee

The gradient term in (\ref{energy}) represents the cost of a
transition between two phases, while the term $F$ represents the local
free energy density for a homogeneous system. We have chosen $F(t) $
to have two symmetric minima corresponding to the case of a
ferromagnetic Ising spin system (and its analogues) below $T_c$.
(Since we are not concerned here explicitly with the dependence on
temperature we have used the scaling parameter $\a$ to 
set the value of the mean field spontaneous
magnetization in $a(t)$, at $T<T_c$, equal to unity.)  The third term
on the right represents the long range
antiferromagnetic type interaction. Note that $v$, which can include terms 
decaying as a power law, is {\it
reflection-positive}, see \cite{FILS1}, summable and completely
monotone, \ie, $(-1)^n v^{(n)}(x)\ge 0$ and $\searrow 0$ as $x\to\io$.
The minimum value of this interaction term occurs when $\phi = 0$. It
competes, therefore, with the first two terms, which are minimized
when $\phi (x) = 1$ or $\phi (x) = -1$ for all $x$. (Note, however, 
that in the absence of the gradient square term the minimizer would be an 
infinitely rapidly oscillating $\phi(x)$, between $-1$ and $+1$; see discussion 
of Kac potential in Section 5.)

To state our results, let us first 
recall some standard notation. $H^1(\RRR)$ is the space of functions 
whose first distributional derivative is in $L^2(\RRR)$ and which $\to 0$ as 
$|x|\to\io$. $H^1_{loc}(\RRR)$ is the space of functions whose derivatives are 
in $L^2([a,b])$ for all intervals $-\io<a<b<+\io$ and 
$H^1_0([a,b])$ are $H^1$ functions that vanish on the endpoints $a,b$. 
In one dimension, $H^1$ functions
are equivalent to H\"older continuous functions, with H\"older exponent $1/2$. 
(See \cite{LL01}). We next define the notions of 
{\it infinite volume specific ground state free energy}
and of {\it infinite volume ground state}. \\

{\bf Definition 1.} {\it Given a finite interval $[a,b]$ on the real line,
let $\EE^{\rm F}_{a,b}: H^1([a,b]) \to \RRR^+$
be the finite volume functional with free boundary conditions, defined as
\be \EE^{\rm F}_{a,b}(\phi)=\int_{a}^{b} dx \Big[\big(\phi'(x)
\big)^2+F(\phi)\Big]+
\int_{a}^{b} dx \int_{a}^{b} dy\; \phi(x) v(x-y) \phi(y)\;.\label{1.2}\ee

Moreover, let $\EE^{\rm D}_{a,b}$ be the restriction of $\EE^{\rm F}_{a,b}$ to 
$H_0^1([a,b])$, that is the finite volume functional with Dirichlet boundary 
conditions $\phi(a)=\phi(b)=0$. Let 
\bea &&E_L^{\rm F}\= \inf_{\phi\in H^1([0,L])}\EE^{\rm F}_{0,L}(\phi)\nn\\
&&E_L^{\rm D}\= \inf_{\phi\in H_0^1([0,L])}\EE^{\rm D}_{0,L}(\phi)\;.\label{1.3}\eea 
Then we define the {\bf infinite volume specific ground state free energy} 
$e_0$ corresponding to the formal energy functional $\EE(\phi)$ to be 
\be e_0=\lim_{L\to\io} E_L^{\rm F}/L=\lim_{L\to\io} E_L^{\rm D}/L
\;,\label{1.4}\ee
whenever the limits on the r.h.s. exist and are equal.}
\\

{\bf Definition 2.} {\it Given $\phi\in H^1_{loc}(\RRR)$, then, for any 
interval $[a,b]$ on the real line, we define:
\bea \EE_{a,b}(\phi)=\int_{a}^{b} dx \Big[\big(\phi'(x)\big)^2+F(\phi)\Big]&+&
\int_{a}^{b}\; dx \int_{a}^{b}\; dy\; \phi(x)\, v(x-y)\, \phi(y) +\label{1.5}\\
&+& 2
\int_{a}^{b}\; dx \int_{\rrrr\setminus[a,b]} dy\; \phi(x)\, v(x-y)\, \phi(y)\;.
\nn\eea

We say that $\phi_0\in H^1_{loc}(\RRR)$ is an {\bf infinite
volume ground state} of $\EE(\phi)$ if 
$\EE_{a,b}(\phi_0)\le \EE_{a,b}(\psi)$, for all intervals $[a,b]$ and 
all functions $\psi$ coinciding with $\phi_0$ on $\RRR\setminus[a,b]$.}
\\

In the following we shall exploit the reflection-positivity of $v$
in order to show existence of periodic minimizers for the
functional (\ref{1.1}). We need to introduce some more definitions
and notation. 

{\bf Definition 3.} {\it Let $M,N\in\ZZZ^+\cup\{+\io\}$ be such that $M+N\ge1$.
Let $\FF=\{f_{-M+1},\ldots,f_0,f_1,\ldots,f_N\}$
be a sequence of functions such that $f_i\in H^1_0([0,T_i])$ and $T_i>0$, with
$-M< i\le N$. 
Let $x_{-M}=-\sum_{j=-M+1}^0 T_j$ and 
$x_i=x_{-M}+\sum_{j=-M+1}^i T_j$, for all $-M< i\le N$ (if $M=0$ it is 
understood that $x_0=0$). Then we define
$\f[\FF]\in H^1_0([x_{-M}, x_N])$
to be the function obtained by juxtaposing the functions $f_i$ on the real 
line, in such a way that, if $x_{i-1}\le x\le x_i$, then 
$\f[\FF](x)=f_i(x-x_{i-1})$, for all $i=-M+1,\ldots,N$.}
\\

In order to visualize the meaning of Definition 3,  we plot, in Figure 1,  a
function $\f[\FF](x)$ corresponding to $M=N=2$.\\

\begin{figure}[ht]
\hspace{1 cm}
\includegraphics[height=4.5truecm]{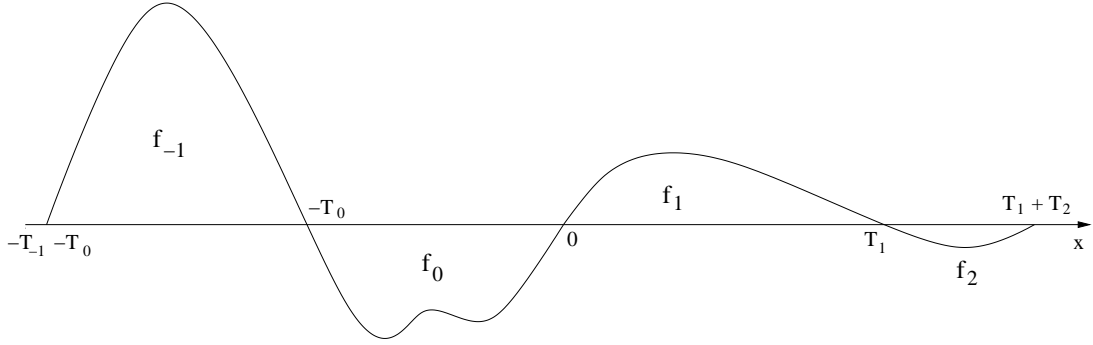}
%\vspace{0.8truecm}
\caption{A possible function $\f[\FF]$ before reflection.}
\end{figure}

{\bf Definition 4.} 
{\it (i) Given $T>0$ and $f\in H^1_0([0,T])$, we define $\th f\in H^1_0([0,T])$
to be the reflection of $f$, namely $\th f(x)=-f(T-x)$, 
for all $x\in[0,T]$.\\
(ii) If $f\in H^1_0([0,T])$, we define $\f[f]=\f[\FF_\infty(f)]\in 
H^1_{\rm loc}(\RRR)$,
where
$\FF_\infty(f)=\{\ldots,f_0,f_1,\ldots\}$ is 
the infinite sequence with $f_n=\th^{n-1} f$.
\\
(iii) Given a sequence $\FF=\{f_{-M+1},\ldots,f_N\}$ as in Def.3,
we define $\FF_-=\{f_{-M+1},\ldots,f_0\}$ and $\FF_+=\{f_1,\ldots,f_N\}$
(if $M=0$ or $N=0$, it is understood that $\FF_-$ or, respectively, $\FF_+$ 
is empty) and we write $\FF=(\FF_-,\FF_+)$.\\
(iv) The reflections of $\FF_-$ and $\FF_+$ are
defined to be: $\th\FF_-=\{\th f_0,\ldots,\th f_{-M+1}\}$ and $\th\FF_+=
\{\th f_N,\ldots,\th f_1\}$.}   See Figure  2.\\

\begin{figure}[ht]
\hspace{3 cm}
\includegraphics[height=8truecm]{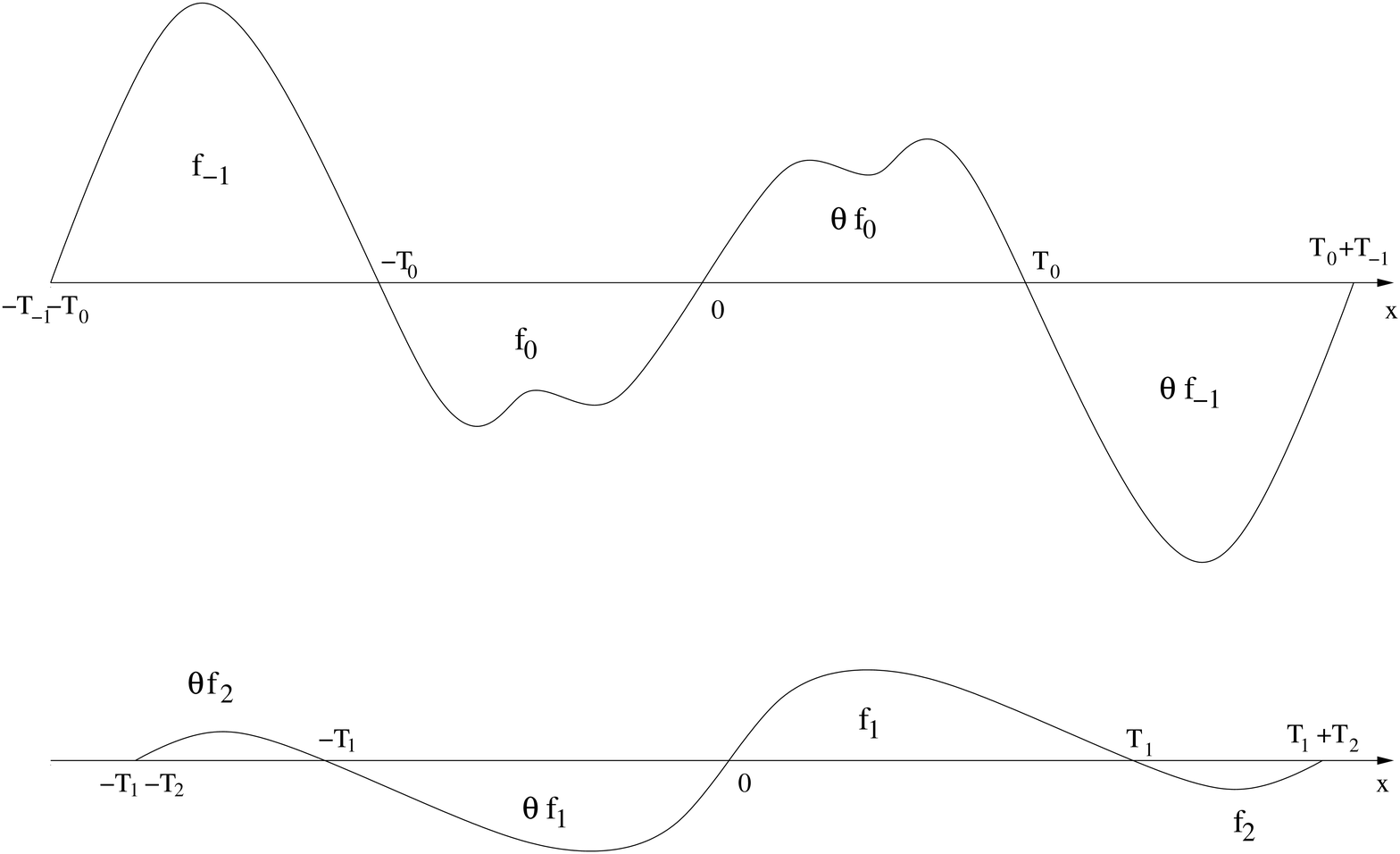}
%\vspace{0.8truecm}
\caption{The two reflected configurations $\f[\FF_1]$ and $\f[\FF_2]$ 
obtained from the function $\f[\FF]$ in Fig.1 after reflection around $0$.}
\end{figure}

We are now ready to state our main results.\\

{\bf Theorem 1. (Specific ground state free energy).}
{\it  For any $T>0$, let $\CC_T=\{f\in H^1_0([0,T]) : 
f\ge 0\}$. The limits in (\ref{1.4}) exist, are equal and are given by
\be e_0=\inf_T e_T\;,\qquad\qquad e_T\=\inf_{f\in \CC_T}
e_\io(f)\;,\label{1.6}\ee
where,
\be e_\io(f)=\lim_{L\to\io} \frac{\EE^{\rm F}_{-L,L}(\f[f])}{2L}\;. \label{1.7}
\ee
(Note that the limit in the r.h.s. of (\ref{1.7}) exists, because $\f[f]$
is periodic and $v$ is summable.)

Moreover $e_T$ is a continuous function of $T$ and 
$\lim_{T\to\io}e_T$ exists and equals $e_0$. It is given by
\be \lim_{T\to\io}e_T=\inf_{f\in\CC_\io}\liminf_{L\to\io} 
\frac{\EE^{\rm F}_{-L,L}(f)}{2L}\;, \label{1.7a}\ee
where 
$\CC_\io=\{f\in H^1_{\rm loc}(\RRR) : 0\le f\le 1\}$.

There is a function, $\phi_T$ that is a minimizer for (\ref{1.6}) and satisfies
$|\phi_T|\le 1$. If $F(t)$ is differentiable for
$t>0$, then $\phi_T$ is twice differentiable and it satisfies the
Euler-Lagrange equation
\be\phi_T''(x)=\frac12 F'(\phi_T(x))+\int_{-\io}^{+\io}dy \; v(x-y)\,
\big(\f[\phi_T]\big)(y)\;.\label{1.7b}\ee
If $F(t)$ is convex for $t>0$, then $\phi_T$ is unique and the inf on the 
r.h.s. of (\ref{1.7a}) is a minimum, with a constant function as a 
minimizer.}\\

{\bf Corollary 1. (Infinite volume ground states).} {\it (i) If there exists 
$T_0$ such that $e_0=e_{T_0}=e_\io(\phi_{T_0})$, then $\f[\phi_{T_0}]$ is an 
infinite volume ground state of $\EE(\phi)$.\\
(ii) If $e_0=\lim_{T\to\io} e_T$ and $F(t)$ is convex for $t\ge 0$, 
the constant function $\phi\=t_0$, with $t_0>0$ the point at which 
$F(t)+t^2 \int_{-\io}^{+\io} v(x)\,dx$ achieves its minimum for $t>0$,
is an infinite volume ground state of $\EE(\phi)$. (Of course
so is $\phi\=-t_0$.)}\\

{\bf Remark.} Theorem 1 and its Corollary may be informally stated 
by saying that all the minimizers of $\EE(\phi)$ are either {\it simply 
periodic}, of finite period $T$, 
with zero average, or of constant sign (and are constant if $F$ is convex 
on $\RRR^+$). By ``simply periodic'' we mean that within a period the minimizer
has only one positive and one negative region, with the negative part obtained
by a reflection from the positive part.

%%%%%%%%%%%%%%%%%%%%%%%%%%%%%%%%%%%%%%%%%%%%%%%%%%%%%%%%%%%%%%%%%%%%%%%%%%%%
\section{Proof of the Main Results} \label{three}
%%%%%%%%%%%%%%%%%%%%%%%%%%%%%%%%%%%%%%%%%%%%%%%%%%%%%%%%%%%%%%%%%%%%%%%%%%%%

As proved in \cite{FILS1}, reflection-positivity of the long range potential 
$v$ implies the following basic estimate.\\

{\bf Lemma 1.} {\it Given a finite sequence of functions 
$\FF=\{f_{-M+1}\ldots,f_0,f_1,\ldots,f_N\}=(\FF_-,\FF_+)$, as in Def.3 and 4, 
we have:
\be \EE^{\rm D}_{x_{-M},x_N}(\f[\FF])\ge \frac12\EE^{\rm D}_{-x_{N},x_N}
(\f[\FF_1])+\frac12\EE^{\rm D}_{x_{-M},-x_{-M}}(\f[\FF_2])\;,\label{1.8}\ee
where $\FF_1=(\th\FF_+,\FF_+)=\{\th f_N,\ldots,\th f_1,f_1,\ldots,f_N\}$ and 
$\FF_2=(\FF_-,\th\FF_-)=
\{f_{-M+1},\ldots,f_0,\th f_0,\ldots,\th f_{-M+1}\}$.}\\

In terms of the function $\f[\FF]$ in Figure 1,
the statement of the Lemma is that the energy of this function is larger than 
the average of the energies of the two reflected configurations in Figure 2.

The key technical ingredient in the proof of Theorem 1 is the 
{\it chessboard estimate}, which is obtained from Lemma 1 by repeatedly 
reflecting around different nodes of the function. A chessboard estimate in the
presence of periodic boundary conditions has appeared many times before in the 
literature, see for instance \cite{FILS1}. Here, however, we will need a 
generalization of it to the case of Dirichlet boundary conditions, 
and we proceed as proposed in the Appendix of \cite{GLL07}.
\\

{\bf Chessboard estimate with Dirichlet boundary conditions.} 
{\it Given a finite sequence of 
functions $\FF=\{f_{1},\ldots,f_N\}$, $N\ge 1$, as in Definition 3, 
with $f_i\in H^1_0([0,T_i])$, we have:
\be \EE^{\rm D}_{0,x_N}\big(\f[\FF]\big)\ge \sum_{i=1}^N T_i e_\io(f_i)
\;.\label{1.9}\ee}

{\cs Proof of (\ref{1.9}).} We proceed by induction.\\
(i) If $N=1$, let us first compare the energy of $f_1$ with that of 
$\{f_1,\pm\, \th f_1\}$. Using the fact that $F(t)$ is even,
the energy of $\{f_1,\pm\, \th f_1\}$ can be rewritten as:
\bea \EE^{\rm D}_{0,2x_1}(\f[\{f_1,\pm\, \th f_1\}])&&=
2\EE^{\rm D}_{0,x_1}(f_1) \pm 2\int_0^{x_1} dx 
\int_{x_1}^{2x_1} f_1(x) v(x-y)\ \th f_1(y-x_1)\=\nn\\
&&\= 2\EE^{\rm D}_{0,x_1}(f_1) +
E_{int} (f_1;\pm \th f_1)\;.\label{1.10}\eea
At least one of the two interaction energies 
$E_{int} (f_1; \th f_1)$ or $E_{int} (f_1; -\th f_1)$ is $\le 0$, simply 
because $E_{int} (f_1; -\th f_1)=-E_{int} (f_1;\th f_1)$. By reflection
positivity, \ie, by Lemma 1, we have in fact that 
$\EE^{\rm D}_{0,2x_1}(\f[\{f_1,\pm\, \th f_1\}])\ge \EE^{\rm D}_{0,2x_1}(\f[\{f_1,
\th f_1\}])$, therefore $E_{int} (f_1; \th f_1)\le 0$. Using (\ref{1.10}) 
we find:
\be \EE^{\rm D}_{0,x_1}(f_1)\ge \frac12\EE^{\rm D}_{0,2x_1}(\f[\{f_1,\th f_1\}])\;.
\label{1.11}\ee
Iterating the same argument, we find:
\be \EE^{\rm D}_{0,x_1}(f_1)\ge \frac{\EE^{\rm D}_{0,2^mx_1}(\f[f_1^{\otimes 2^m}])}{2^m}
\;,\label{1.12}\ee
where, by definition, 
\be f_1^{\otimes 2^m}=\{\,\overbrace{f_1,\th f_1,\ldots,f_1,\th f_1}^{2^m\ 
{\rm times}}\,\}\;.\label{1.13}\ee
Taking the limit $m\to\io$ in (\ref{1.12}) we get the desired estimate: 
\be\EE^{\rm D}_{0,x_1}(f_1) \ge T_1 e_\io(f_1)\;.\label{1.13a}\ee
(ii) Let us now assume by induction that the bound is valid for all 
$1\le N\le n-1$,
$n\ge 2$, and let us prove it for $N=n$. There are two cases.

\\(a) $n=2p$ for some $p\ge 1$. If we reflect once, by Lemma 1 we have:
\bea \EE^{\rm D}_{0,x_{2p}}(\f[\{f_1,\ldots,f_{2p}\}])&\ge&
\frac12
\EE^{\rm D}_{0,2(x_{2p}-x_p)}(\f[\{\th f_{2p},\ldots,\th f_{p+2}, 
(\th f_{p+1})^{\otimes 2}, f_{p+2},\ldots f_{2p}\}])+\nn\\
&+& \frac12 
\EE^{\rm D}_{0,2x_p}(\f[\{
f_1,\ldots,f_{p-1},f_p^{\otimes 2},\th f_{p-1},\ldots,\th f_1\}]) 
\label{1.14}\eea
If we now regard $(\th f_{p+1})^{\otimes 2}$ and $f_p^{\otimes 2}$
as two new functions in $H^1_0([0,2T_{p+1}])$ and in $H^1_0([0,2T_{p}])$,
respectively, the two terms in the r.h.s. of (\ref{1.14}) can be regarded as 
two terms with $N=2p-1$ and, by the induction assumption, they satisfy the 
bounds:
\bea && \EE^{\rm D}_{0,2(x_{2p}-x_p)}
(\f[\{\th f_{2p},\ldots,\th f_{p+2}, (\th f_{p+1})^{\otimes 2},f_{p+2},\ldots 
f_{2p}\}])\ge 2 \sum_{i=p+1}^{2p} T_i e_\io(f_i)\;,\nn \\
&& \EE^{\rm D}_{0,2x_p}(\f[\{f_1,\ldots,f_{p-1},f_p^{\otimes 2},\th f_{p-1},
\ldots,\th f_1\}])\ge 
2 \sum_{i=1}^{p} T_i e_\io(f_i)\;, \label{1.15}\eea
where we used that $e_\io((\th f_{p+1})^{\otimes 2})=e_\io(f_{p+1})$ and 
$e_\io(f_p^{\otimes 2})=e_\io(f_p)$. Therefore, the desired bound is proved. 

\\(b) $n=2p+1$ for some $p\ge 1$. If we reflect once we get: 
\bea && \EE^{\rm D}_{0,x_{2p+1}}(\f[\{f_1,\ldots,f_{2p+1}\}])\ge\label{1.16} \\
&& \ge \frac12 
\EE^{\rm D}_{0,2(x_{2p+1}-x_{p+1})}(\f[\{\th f_{2p+1},\ldots,\th f_{p+3},
(\th f_{p+2})^{\otimes 2},f_{p+3},\ldots,f_{2p+1}\}])+\nn\\ 
&& \hspace{.1truecm}+\frac12 
\EE^{\rm D}_{0,2x_{p+1}}(\f[\{f_1,\ldots,f_p,f_{p+1}^{\otimes 2},\th f_p,\ldots,
\th f_1\}]) \nn\eea
The first term in the r.h.s. corresponds to $N=2p-1$ so by the induction 
hypothesis it is bounded below by $\sum_{i=p+2}^{2p+1} T_i 
e_\io(f_i)$. As regards the second term, using reflection positivity again, 
we can bound it from below by
\be \frac14
\EE^{\rm D}_{0,2x_p}(\f[\{f_1,\ldots,f_{p},\th f_{p},\ldots,\th f_1\}])+\frac14 
\EE^{\rm D}_{0,2x_{p}+4x_{p+1}}(\f[\{f_1,\ldots,f_{p},(f_{p+1})^{\otimes 4},
\th f_{p},\ldots,\th f_1\}])\label{1.17}\ee
By the induction hypothesis, the first term is bounded below by 
$(1/2)\sum_{i=1}^p T_i e_\io(f_i)$, and the second can be bounded
using reflection positivity again. Iterating we find:
\bea && \EE^{\rm D}(\f[\{f_1,\ldots,f_{2p+1}\}])\ge\label{1.18} \\
&& \quad \ge \sum_{i=p+2}^{2p+1} T_i e_\io(f_i)
+ \Big(\sum_{n\ge 1}2^{-n}\Big)\cdot \sum_{i=1}^p T_i e_\io(f_i)\Big)
+\nn\\
&&\hspace{2.truecm} +\lim_{n\to\io} 2^{-n} 
\EE^{\rm D}_{0,2x_{p}+2^m x_{p+1}}(\f[\{f_1,\ldots,f_{p},(f_{p+1})^{\otimes 2^m},
\th f_{p},\ldots,\th f_1\}])\;. \nn\eea
Note that the last term is equal to $T_{p+1} e_\io(f_{p+1})$, so (\ref{1.18})
is the desired bound. This concludes the proof of (\ref{1.9}). \qed

\vskip.5truecm

{\cs Proof of Theorem 1.}
(1) Let us first prove that $e_0=\inf_T\inf_{f\in \CC_T} e_\io(f)$. 
First of all, let us note that $\limsup_{L\to\io}E_L^{\rm F}/L=
\limsup_{L\to\io}E_L^{\rm D}/L$
and $\liminf_{L\to\io}E_L^{\rm F}/L=\liminf_{L\to\io}E_L^{\rm D}/L$, because the 
interaction $v$ is absolutely summable. Moreover,
it follows by the variational estimate 
$E_L^{\rm F}\le \EE^{\rm F}_{0,L}(\f[f])$, valid for any $f\in\CC_T$, $T>0$,
that $\limsup_{L\to\io}E_L^{\rm F}/L\le \inf_T\inf_{f\in \CC_T} e_\io(f)$.

We then just need to prove that 
$\liminf_{L\to\io}E_L^{\rm D}/L\ge \inf_T\inf_{f\in \CC_T} e_\io(f)$. 
For this purpose,
given $L>0$, let $\phi$ be any function in $H^1_0([0,L])$ and let us denote by 
$x_0=0<x_1<\ldots<x_N=L$ its nodes (if $\phi$ is identically $0$ in some 
interval $[a,b]\subseteq[0,L]$, it will be understood that $\phi$ has two nodes
between $a$ and $b$, the first located at $x=a$ the second at $x=b$). 
We define: $T_i=x_i-x_{i-1}$, $i=1,\ldots,N$. Let $f_i:
[0,T_i]\to\RRR$ be such that $\phi(x)=f_i(x-x_{i-1})$, for all 
$x\in[x_{i-1},x_i]$. By construction any $f_i$ is either nonnegative or 
nonpositive and $\phi=\f[\{f_1,\ldots,f_N\}]$. Using the chessboard estimate
we find:
\be \EE^{\rm D}_{0,L}(\phi)=\EE^{\rm D}_{0,L}(\f[\{f_1,\ldots,f_N\}])\ge 
\sum_{i=1}^N
T_i e_\io(f_i)\;.\label{1.19}\ee
All the $e_\io(f_i)$ in the r.h.s. can be bounded below by 
$\inf_T\inf_{f\in \CC_T}e_\io(f)$, and the proof of this first claim is 
concluded.\\
\\
(2) Next, let us show that for any fixed $T>0$ there exists a function 
$\phi_T\in\CC_T$ such that $|\phi_T|\le 1$ and $e_\io(\phi_T)=
\inf_{f\in \CC_T}e_\io(f)\=e_T$. Note that, for any $f\in\CC_T$, $e_\io(f)$ can
be rewritten as:
\be e_\io(f)=\frac1T\int_{0}^{T} dx \big(f'(x)\big)^2+ \frac1T \int_{0}^{T} 
dx \, F(f(x)) + \frac1T
\int_{0}^{T} dx \int_{0}^{T} dy\; f(x)\, f(y)\,\widetilde v_T(x,y)\;,
\label{1.20}\ee
with 
\be \widetilde v_T(x,y) = \sum_{n\in\zzz}\Big[
v(2nT+y-x)-v(y+x+2nT)\Big]\;.\label{1.20aa}\ee
Since $F\ge 0$, the first two terms on the r.h.s. of (\ref{1.20})
are clearly nonnegative. Since $v$ is absolutely summable, the potential 
$\widetilde v_T(x,y)$ can be rewritten as
\be \widetilde v_T(x,y)=\sum_{n\ge 0}\Big[v(2nT+y-x)-v\big(2(n+1)T-y-x\big)
-v(2nT+y+x)+v\big(2(n+1)T-y+x\big)\Big]\label{1.20a}\ee
and, using the fact that $v''\ge 0$, it is straightforward to check that each 
term in the sum on the r.h.s. is pointwise positive, for all $0\le x,y\le T$. 
This implies that the third term in the r.h.s. of (\ref{1.20}) is nonnegative
as well. We will denote the kinetic energy, \ie, the first term in the r.h.s. 
of (\ref{1.20}), by $K_f$, and the second and third terms by $V_f$ and $W_f$, 
respectively. 

Now let $f^j$ be a minimizing sequence, \ie, $e_\io(f^j)\to e_T$ as $j\to\io$ 
and $f^j\in\CC_T$. First we note that $K_{f^j}$ is bounded by a constant 
independent of $j$, because $K_{f^j}\le T e_\io(f^j)$ and $e_\io(f^j)$ is 
uniformly bounded from above by some constant $C$. Moreover, we can assume 
without loss of generality that $|f^j|\le 1$: in fact, since  
$F(t)$ has an absolute minimum at $t=1$ and the potential 
$\widetilde v(x,y)$ is pointwise positive, we see that the energy 
in (\ref{1.20}) decreases by replacing $f$ with $\min\{f,1\}$. 
In fact, each of the three terms in the energy can only decrease with the 
replacement.
Then the sequence $f^j$ is bounded in $H^1_0([0,T])$. Since bounded sets in 
$H^1_0([0,T])$ are weakly sequentially compact (see \cite{LL01}, Section 7.18),
we can therefore find a function $\phi_T$ in $H^1_0([0,T])$ and a subsequence
(which we continue to denote by $f^j$) such that $f^j\to \phi_T$ weakly in 
$H^1_0([0,T])$. By Corollary 8.7 in \cite{LL01} (``weak convergence implies 
a.e. convergence''), we can assume without loss of generality that
$f^j(x)\to \phi_T(x)$ for almost every $x\in[0,T]$. 
This function $\phi_T$ satisfies $|\phi_T|\le 1$ and will be our minimizer: 
in fact, since the kinetic energy $K_f$ is weakly lower semicontinuous (see 
\cite{LL01}, Section 8.2), and since 
$V_{f^j}\to V_{\phi_T}$ and $W_{f^j}\to W_{\phi_T}$ 
as $j\to\io$, by the dominated convergence theorem, we have that 
\be e_T=\lim_{j\to\io}e_\io(f_j)\ge e_\io(\phi_T)\label{1.21}\ee
and this shows that $\phi_T$ is the minimizer. 
\\
\\
(3) Now, let us show that $e_T$ is continuous in $T$. We shall do this
by deriving bounds from above and below on $e_{T+\e}$, tending to $e_T$ as
$\e\to 0$. Let us take $\e>0$. In order to get the bound from above on
$e_{T+\e}$, let us consider a variational function $f_{T+\e}
\in H^1_0([0,T+\e])$, coinciding with $\phi_T$, \ie,
the minimizer of $e_T$, for $x\in[0,T]$, and equal to $0$ in $x\in[T,T+\e]$. 
Using (\ref{1.20}) we get
\bea e_{T+\e}\le e_\io(f_{T+\e})&=&
\frac1{T+\e}\int_{0}^{T} dx \big(\phi_T'(x)\big)^2+ \frac1{T+\e}\int_{0}^{T} 
dx \, F(\phi_T(x)) +\frac\e{T+\e}F(0)+\nn\\
&+& \frac1{T+\e}\int_{0}^{T} dx \int_{0}^{T} dy\; \phi_T(x)\, \phi_T(y)\,
\widetilde v_{T+\e}(x,y)\;,\label{1.22}\eea
and clearly $e_\io(f_{T+\e})\to e_\io(\phi_T)$ as $\e\to 0$. 
In order to get a lower bound, let us
use the variational estimate $e_T\le e_\io(g_T)$, where 
$g_T=\phi_{T+\e}\big(x(1+\e/T)\big)$. Using (\ref{1.20}), we get
\bea &&e_T\le e_\io(g_T)=\frac1T\Big(1+\frac\e{T}\Big)\int_{0}^{T+\e} dx 
\big[\phi_{T+\e}'(x)\big]^2+\frac1T\Big(1+\frac\e{T}\Big)^{-1}
\int_{0}^{T+\e} dx \, F(\phi_{T+\e}(x)) +\nn\\
&&\hspace{1.9truecm}
+\frac1T\Big(1+\frac\e{T}\Big)^{-2}\int_{0}^{T+\e} dx \int_{0}^{T+\e} dy\; 
\phi_{T+\e}(x)\,\phi_{T+\e}(y) \,\times\label{1.23}\\
&&\hspace{3.truecm}\times\,
\sum_{n\in\zzz}\Big[v\Big(\frac{2n(T+\e)+y-x}{1+\e/T}\Big)-v\Big(
\frac{2n(T+\e)+y+x}{1+\e/T}\Big)\Big]\nn\eea
and clearly $e_\io(g_T)-e_\io(\phi_{T+\e})\to 0$ as $\e\to 0$. 
A similar proof applies to the case $\e<0$, therefore 
continuity of $e_T$ is proved.\\
\\
(4) Let us prove that $\lim_{T\to\io}e_T$ exists and is equal to:
\be \lim_{T\to\io}e_T=\inf_{f\in\CC_\io} \liminf_{L\to\io} 
\frac{\EE^{\rm F}_{-L,L}(f)}{2L} \;,\label{1.25}\ee

where $\CC_\io=\{f\in H^1_{loc}(\RRR) : 0\le f\le 1\}$. For this purpose, if
$\phi_T$ is the minimizer of $e_T$, let us rewrite
\bea &&e_\io(\phi_T)=\frac1T\int_{0}^{T} dx \Big[\big(\phi_T'(x)\big)^2+ 
F(\phi_T(x))\Big]+\frac1{T} \int_{0}^{T} dx \int_{0}^{T} dy\; 
\phi_T(x)\, \phi_T(y)\, v(y-x)+\nn\\
&&+\frac1{T} \int_{0}^{T} dx \int_{0}^{T} dy\; \phi_T(x)\, \phi_T(y)\, 
\sum_{n\ge 1}\Big[v(2nT+y-x)-v(2nT-y-x)\Big]+\label{1.26}\\
&&+\frac1{T} \int_{0}^{T} dx \int_{0}^{T} dy\; \phi_T(x)\, \phi_T(y)\, 
\sum_{n\ge 1}\Big[v(2nT-y+x)-v(2(n-1)T+y+x)\Big]\;.\nn\eea
Using the fact that $0\le \phi_T\le 1$, 
as proved in part (2), and the fact that $v\in L^1(\RRR)\cap L^\io(\RRR)$ 
is completely monotone, we find that the last two terms in (\ref{1.26})
tend to $0$ as $T\to\io$. Therefore:
\be e_T=\frac{\EE^{\rm D}_{0,T}(\phi_T)}T+o(1)= \inf_{f\in \CC^*_T}
\frac{\EE^{\rm F}_{0,T}(f)}T+o(1)\label{1.27}
\;,\ee
where $\CC^*_T=\{f\in H^1([0,T]) : 0\le f\le 1\}$. Repeating the proof in part
(2), we find that the inf in the r.h.s. is a minimum, and we denote by $f_T$ 
the corresponding minimizer. Note that the quantity
$\EE^{\rm F}_{0,T}(f_T)$ is superadditive in $T$, \ie, 
$\EE^{\rm F}_{0,T_1+T_2}(f_{T_1+T_2})\ge \EE^{\rm F}_{0,T_1}(f_{T_1})
+\EE^{\rm F}_{0,T_2}(f_{T_2})$. 
Then the limit as $T\to\io$ of 
$\EE^{\rm F}_{0,T}(f_T)/T$ exists and
\be \lim_{T\to\io}e_T=\lim_{T\to\io}\frac{\EE^{\rm F}_{0,T}(f_T)}T\;.\label{1.28}\ee
Now, on the one hand, using the variational estimate 
$\EE^{\rm F}_{0,T}(f_T)\le \EE^{\rm F}_{0,T}(f)$ 
valid for any $f\in\CC_\io$, we see
that the limit on the r.h.s. of (\ref{1.28}) is smaller than $\inf_{f
\in\CC_\io} \liminf_{L\to\io} \EE^{\rm F}_{-L,L}(f)/(2L)$. On the other hand,
using summability and complete monotonicity of $v$, we find that for any $T$
\be \frac{\EE^{\rm F}_{0,T}(f_T)}T=\lim_{L\to\io}\frac{\EE^{\rm F}_{-L,L}
(\widetilde\f[f_T])}{2L}+o(1)\;,\label{1.29}\ee
where $\widetilde\f([f_T])\in\CC_\io$ is the function obtained by periodically
repeating the sequence $\{f_T,-\th f_T\}$ infinitely many times and $o(1)$ is
a remainder that goes to $0$ as $T\to\io$. Clearly, the first term in the 
r.h.s. can be bounded from below by $\inf_{f\in\CC_\io} \liminf_{L\to\io} 
\EE^{\rm F}_{-L,L}(f)/(2L)$ and this concludes the proof of the claim.
\\
\\
(5) Finally, it is straightforward to check that if the distributional 
derivative of $F$ is a function, then the minimizer $\phi_T$ satisfies the 
Euler-Lagrange equation in the sense of distributions. If $F(t)$ is 
differentiable for $t>0$, by the smoothness of $v$, it follows by a standard 
``bootstrap'' argument (see Theorem 11.7 in \cite{LL01}), that $\phi_T\in C^2$.

If $F(t)$ is convex for $t>0$, then the functional $e_\io(f_T)$ is strictly 
convex (because $\int (f')^2$ is strictly convex)
for $f_T\in \CC_T$ and the minimizer $\phi_T$ is unique. Similarly,
for any $L>0$, the functional $\EE^{\rm F}_{0,L}(f)/L$ is strictly convex 
for $f\in\CC^*_L$ and so is $\EE^{\rm per}_{0,L}(f)/L$, where 
$\EE^{\rm per}_{0,L}(f)$ is the analogue of $\EE^{\rm F}_{0,L}(f)$ with periodic 
boundary conditions:
\be \EE^{\rm per}_{0,L}(f)=\int_{0}^{L} dx\Big[\big(f'(x)\big)^2+F(f)\Big]+
\int_{0}^{L} dx \int_{0}^{L} dy f(x) f(y) \sum_{n\in\zzz} v(nL+y-x) 
\;,\label{1.29a}\ee
with $f\in\CC^{per}_L\=\{g\in \CC^*_L\ :\ g(0)=g(L)\}$. 

By the summability of $v$, 
\be \liminf_{L\to\io} 
\frac{\EE^{\rm F}_{0,L}(f)}{L} =\liminf_{L\to\io} 
\frac{\EE^{\rm per}_{0,L}(f)}{L}\;,\qquad \forall f\in\CC_\io\;.
\label{1.30}\ee
By periodicity, $\EE^{\rm per}_{0,L}(f)=\frac1{L}\int_{0}^L 
\EE^{\rm per}_{0,L}
(\t_x f)$, where $\t_x f(y)\=f(y-x)$. By convexity, the latter quantity is 
bounded below by $\EE^{\rm per}_{0,L}(\media{f})=L\big[F(\media{f})
+\media{f}^2\int_{0}^{L}\frac{d x}L\int_{0}^{L}d y\,v(x-y)\big]$. 
This shows that the limit 
as $T\to\io$ of $e_T$ is $\min_{t\in\rrr^+}\{F(t)
+t^2\int_{-\io}^{+\io}d x\,v(x)\}$ and concludes the proof of Theorem 1.
\qed

Corollary 1 is a simple consequence of Theorem 1 and of its proof.\\

{\cs Proof of Corollary 1.} {\it (i)}. Let $T_0$ be such that $e_0=e_{T_0}$
and let us assume by contradiction that there exists an interval $[a,b]$  
and a function $f\in H^1_{loc}(\RRR)$, coinciding with $\f[\phi_{T_0}]$ 
on $\RRR\setminus[a,b]$, and such that $\EE_{a,b}(f)<\EE_{a,b}
(\f[\phi_{T_0}])$. Note that, if $[a',b']\supseteq[a,b]$, then 
$\EE_{a',b'}(f)-\EE_{a',b'}(\f[\phi_{T_0}])=\EE_{a,b}(f)-
\EE_{a,b}(\f[\phi_{T_0}])$. We choose $[a',b']\supseteq[a,b]$ such that
$b'-a'=k T_0$, for some $k\in \NNN$, and $f(a')=f(b')=0$. 
We denote by $f_1$ the restriction of $f$ to $[a',b']$ and we write:
\bea&& 0> \EE_{a,b}(f)-\EE_{a,b}(\f[\phi_{T_0}])\label{1.31}\\
&&\qquad=
\lim_{m\to\io}\Big[\EE^{\rm D}_{a'-mT_0,b'+mT_0}\big(\f[\{\phi_{T_0}^{\otimes m},
f_1,\phi_{T_0}^{\otimes m}\}]\big)-\EE^{\rm D}_{a'-mT_0,b'+mT_0}\big(\f[\phi_{T_0}]
\big)\Big] \nn\\
&&\qquad
=\lim_{m\to\io}\Big[\EE^{\rm per}_{a'-mT_0,b'+mT_0}
\big(\f[\{\phi_{T_0}^{\otimes m},f_1,\phi_{T_0}^{\otimes m}\}]\big)-
\EE^{\rm per}_{a'-mT_0,b'+mT_0}\big(\f[\phi_{T_0}]\big)\Big]\;.\nn\eea
Now, $\EE^{\rm per}_{a'-mT_0,b'+mT_0}\big(\f[\phi_{T_0}]\big)=(2m+k)T_0 e_0$
and, by the chessboard inequality, 
\be \EE^{\rm per}_{a'-mT_0,b'+mT_0}\big(\f[\{\phi_{T_0}^{\otimes m},f_1,
\phi_{T_0}^{\otimes m}\}]\big)\ge 2mT_0 e_0+k T_0 e_\io(f_1)\;,\label{1.32}\ee
so that we find $e_\io(f_1)-e_0<0$, which is a contradiction.\\
\\
{\it (ii)} Let $e_0=\lim_{T\to\io}e_T$ and $F(t)$ convex for $t>0$. 
As proved in Theorem 1, $e_0=\min_{t\in\rrr^+}\{F(t)
+t^2\int_{-\io}^{+\io}d x\,v(x)\}\= F(t_0)+t_0^2 \int_{-\io}^{+\io}d x\,v(x)$
and a repetition of the proof in part (i) shows that $f(x)\= t_0$
is an infinite volume ground state. \qed

%%%%%%%%%%%%%%%%%%%%%%%%%%%%%%%%%%%%%%%%%%%%%%%%%%%%%%%%%%%%%%%%%%%%%%%%%%%%%%
\section{An example}\label{four}
%%%%%%%%%%%%%%%%%%%%%%%%%%%%%%%%%%%%%%%%%%%%%%%%%%%%%%%%%%%%%%%%%%%%%%%%%%%%%%

One expects that when $v(x)x$ is summable, i.e., $\int_0^\io \n(d\a) 
\a^{-2}<+\io$, the minimizer is periodic when $\l=v(0)$ is large, while small
$\lambda$ produces a function with constant sign, say positive. If
$v(x)x$ is not summable one expects that the minimizer is always
periodic. We shall not prove this last statement, but see \cite{GLL06}
for a similar discussion in the discrete microscopic case.

Here we give an illustrative example that will make this small $\lambda$,
large $\lambda $ dichotomy clear. 
This example is generic, in the convex case, at least,
it is only a question of estimating orders of magnitude in the two
regimes of $\l$.

Let $v(x) =\l e^{-|x|}$ and $ F(\phi)=(|\phi| -1)^2$. This is the
``convex case'', in the sense that $F''(\phi)>0$ for $\phi>0$. 
When $\lambda =0$ the minimum energy occurs when $\phi'
=0$ and $F=0$, which means that either $\phi(x) =1$ for all $x$
or $\phi=-1$ for all $x$. 
For small
$\lambda$, $\phi$ will be of constant sign, and hence a constant, by
convexity. 
To see this it suffices to note  that $|\phi(x)|$ must be nearly
$1$ for most $x$ (by continuity of the energy), and if $\phi$ had a
jump from $+1$ to $-1$ the cost in kinetic energy $\int |\phi'|^2$
would outweigh any gain in the integral term coming from the
interaction of a negative $\phi$ region and a positive $\phi$ region --
which would be of the order of $\lambda$, at best.

To show that one gets periodicity for large $\lambda$ it is only
necessary to write down the energy for the constant $\phi $ case and
compare it with a crude variational periodic $\phi$. The constant is easily
calculated to be $\phi_0 = (1+2\lambda)^{-1} \ll 1$ and the specific
energy is $e= 2\lambda (1+2\lambda)^{-1}$. The variational function
can be taken to be $\pm \phi_0 $ with a large period $T$ and with a
linear interpolation between $+\phi_0$ and $-\phi_0$ of width 
$\beta\sim\l^{-1/2}\ll 1$.  
This gives a local energy (\ie, the first term in (\ref{1.1})) $\sim
\l^{-3/2}$ for each such interface.  
The gain in interaction energy across the interface is $\sim -\l^{-1}$,
which is greater than this.

There is no need to belabor the details of such examples. The
conclusion is that there must be a transition from constant to
periodic as $\lambda$ increases. The critical $\lambda_c$ at which the 
transition occurs can be computed by imposing the condition that the energy of 
the ``kink'', \ie, the antisymmetric solution to the Euler-Lagrange equation 
with boundary conditions $\phi(\pm \infty) =
\pm \phi_0$ and $\phi(0) = 0$, is the same as that of the constant function
$\phi(x)\=\phi_0$ (note that both energies are infinite, but the energy 
difference is well defined and finite). In our example
the kink solution $\phi$ can be computed {\it exactly}, and likewise its
energy. To be specific, let us write the Euler-Lagrange equation 
for $\phi(x)$, $x\ge 0$, as:
\be -\phi''(x)+\phi(x)-1+\l\int_0^\io dy\, 
\big(e^{-|x-y|}-e^{-x-y}\big)\,\phi(y)=0\;.\label{2.1}\ee
Defining $h(x)=\int_0^\io dy\, e^{-|x-y|} \phi(y)$ and $c=\int_0^\io dy\, 
e^{-y}\phi(y)$, we can rewrite this equation as
\be -\phi''(x)+\phi(x)-1+\l h(x) -\l c e^{-x}=0\;.\label{2.2}\ee
This implies that $\phi''(0)=-1$. If we differentiate twice and use
the fact that $h''(x)=h(x)-2\phi(x)$, we find:
\bea && -\phi''''(x)+\phi''(x)+\l h(x) -2\l \phi(x)-\l c e^{-x}=\nn\\
=&&-\phi''''(x)+2\phi''(x)-(1+2\l)\phi(x)+1=0\;,\label{2.3}\eea
where we used (\ref{2.2}) to rewrite $\l h(x) -\l c e^{-x}=
\phi''(x)-\phi(x)+1$. The only solution to (\ref{2.3}) satisfying $\phi(0)=0$,
$\phi(+\io)=\phi_0=(1+2\l)^{-1}$ and $\phi''(0)=-1$ is 
\be \phi(x)= \frac1{1+2\l}\Big(1-e^{-\m_1x}\frac{\cos(\m_2x+\th)}{\cos\th}
\Big)\;,\label{2.4}\ee
where:
\be \m=\m_1+i\m_2=
(1+2\l)^{1/4} e^{i\th/2}\qquad{\rm and}\qquad\th=\arcsin
\sqrt{\frac{2\l}{1+2\l}}\;.\label{2.5}\ee
The difference between the energy of the solution in (\ref{2.4}) and that
of the constant function $\phi(x)=(1+2\l)^{-1}$ is
\be \D E=\frac2{1+2\l}
\int_0^\io dx\, e^{-\m_1x}\frac{\cos(\m_2x+\th)}{\cos\th}
\label{2.6}\ee
and imposing $\D E=0$ we get the condition $\cos(3\th/2)=0$, which 
is equivalent to 
\be \l=\l_c=\frac32\;.\ee
The conclusion is that in our example the infinite volume ground state is 
the constant function $\phi(x)=(1+2\l)^{-1}$, for all $\l\le 3/2$, and 
is periodic, for all $\l>3/2$. 
\bigskip

\section{Discussion and connection 
with related work}\label{five}

We investigated a class of 1D
free-energy functionals, characterized by a competition between a
local term, prefering a constant minimizer (equal to $1$ or $-1$, the 
positions of the minima of
a ``double-well'' even function $F$), and a long range positive
interaction, which is assumed to be reflection positive and summable.
We showed by reflection positivity that, for any strength of the long
range interaction, the ground state is either periodic (with mean
zero) or of constant sign. If the local term $F(\phi)$, besides being even in
$\phi$, is assumed to be convex on $\RRR^+$, then the ground state is
either periodic or constant. The proof is simple and does not
depend on the details of the function $F$ or $v$ (as long as $v$ is
positive and reflection positive). Note, however, that the assumption
that $F$ is even is crucial: this means that we cannot include a
chemical potential different from zero.

Our results extend or complement some known results first proved by Alberti and
M\"uller \cite{AM} and by Chen and Oshita \cite {CO} on periodicity 
of the minimizers of certain 1D free-energy functionals.
They consider models similar to (\ref{1.1}), with
non-zero chemical potential. They require however a smallness condition
on the gradient term and there are certain classes of potentials, e.g., 
power law type interactions, which are not included in their proofs. Their 
analysis is based 
on detailed apriori asymptotic estimates that are not needed in our approach. 

Let us conclude by mentioning the connection of our results with the so-called 
``froth problem'' put forward by Lebowitz and Penrose in \cite{LP}.
They consider $d$-dimensional systems of particles 
with density $\r$ (or spin systems with magnetization $m$)
interacting both with a short range interaction and with a long range Kac 
potential of the form $\g^d v(\g{\bf r})$, with $\int_{\rrr^d}v({\bf r})d{\bf r}
=\a$. When
$\g\to 0$, the exact free energy per unit volume, $a(\r)$, for the case where 
$v$ is positive definite (which includes the cases considered here) is 
given by $a(\r)=a_s(\r)+\frac12\a\r^2$; $a_s(\r)$ is the free energy due 
to the short range potential. In cases where the short range interaction 
induces a phase separation, as indicated by a linear segment in $a_s(\r)$, 
the long range positive definite Kac potential, with $\a>0$, will
lead to a strictly convex $a(\r)$. This means that 
the global phase segregation, due to the short range interaction,
is destroyed by the long range positive definite Kac potential in the
limit $\g\to0$.
The interpretation given in \cite{LP} was
that there is no phase separation on the scale $\g^{-1}$, but 
non trivial structures may appear on an intermediate scale 
$1\ll\g^{-\d}\ll\g^{-1}$.
In this sense the system for finite, but small, $\g$ is expected to be a sort 
of froth, with structures invisible on large scales, but observable on 
intermediate scales
(and these structures may form periodic patterns, as discussed
in the introduction). While the scale $\g^{-\d}$ is unknown in general,
our results on microscopic models show that, at least in 1D lattice models, 
the correct scale to look at, at zero temperature, is $\g^{-2/3}$, see 
\cite{GLL06} and the discussion in Section VIII of \cite{GLL07}. 

The problem of understanding these mesoscopic structures
can be related to the minimization problem studied in this paper,
thanks to a result by Gates and Penrose \cite{GP}, who proved that 
the free energy $a(\r)$ can also be obtained from a minimization 
of a free energy functional similar to (\ref{1.1}), but 
without a gradient term. Such functional has the interpretation of 
large deviation functional for observing a mesoscopic density \cite{BBP,DE,
BBBP,CCELM}. As already noted, the absence of a gradient term 
in this functional means that its minimizers (the ``typical mesoscopic 
configurations'') will oscillate on a scale small 
compared to $\g^{-1}$, e.g., the $\g^{-2/3}$ found in \cite{GLL07}. 
It would be nice to understand the correspondence between these oscillations 
for $\g\ll1$ and the ones found here and in \cite{AM}.

\acknowledgments 

We thank Paolo Butt\`a, Anna De Masi, Errico Presutti and particularly 
Eric Carlen for valuable discussions.  
The work of JLL was supported by NSF Grant DMR-044-2066 and by AFOSR
Grant AF-FA 9550-04-4-22910. The work of AG and EHL 
was partially supported by U.S. National Science Foundation
grant PHY-0652854.

\end{document}